# Spin Noise Fluctuations from Paramagnetic Molecular Adsorbates on Surfaces

Paolo Messina,[*] Matteo Mannini, Andrea Caneschi, Dante Gatteschi, Lorenzo Sorace
*Department of Chemistry and INSTM Research Unit , University of Florence, 50019, Sesto Fiorentino, Italy*
Paolo Sigalotti, Cristian Sandrin,
*APEResearch, Science Park Area, campus Basovizza, 34012, Trieste, Italy*
Paolo Pittana
*Sincrotrone, Science Park Area, campus Basovizza, 34012, Trieste, Italy*
Yishay Manassen
*Department of Physics, Ben Gurion University of the Negev, P.O.B 653 Beer Sheva 84105, Israel*

## ABSTRACT

The measurement of spin noise in nuclei was pioneered on bulk samples more than two decades ago. An ensemble of spins can produce a coherent signal at the frequency of a static magnetic field, known as spin noise, an effect due to the statistical polarization of small ensembles. The difficulty of these measurements is that the signal is extremely small – even if electron spins are detected. Although the statistical polarization of N spins dominates the Boltzmann statistics if N approaches unity, a more sensitive tool is requested to measure the polarization of the magnetic moment of a single spin. In this paper we report on the verification of recent results on the detection of spin noise from paramagnetic molecules of BDPA (α,γ-Bisdiphenylene-β-phenylallyl) by Durkan and coworkers.[C. Durkan and M. E. Welland, *Appl. Phys. Lett*. **80**, 458 (2002)] We also present new results on a second paramagnetic specie, DPPH (1,1-Diphenyl-2-picrylhydrazyl),  deposited on Au(111) surfaces.  ESR spectra from ultrathin films of DPPH and BDPA  grown on Au(111) are reported. We prove that the paramagnetic molecules preserve their magnetism on the surface. These data and a thorough analysis of the signal recovery apparatus help to understand the low statistical recurrence of the spin noise in the data set . A thorough description of the experimental apparatus together with an analysis of the parameters that determine the sensitivity are also presented.

PACS numbers:

## I. INTRODUCTION

An ensemble of $N$ spins of magnetic moment $\mu$ can produce a statistical polarization of its magnetization proportional to $N^{1/2}\mu$ without the application of any driving Radio Frequency (RF) field. These measurements have received much attention since the detection of spin noise from $^{35}Cl$ as they represent one of the fundamental issues in magnetic resonances.[1-3] In fact the detection of magnetic resonances through the detection of the statistical polarization of the magnetic moments can open up the possibility to detect the spin dynamics of small ensembles.[3] In these systems a conventional probing apparatus might disruptively interact with the system under investigation. There is therefore a need for new probes that do not alter the state of the physical system under investigation. In recent experiments[3] the detection of spin noise has led to determine $g$-factor, electron spin, nuclear spin, hyperfine splittings, nuclear moments and spin coherence life time of a small ensemble of electron spins.

Coherence effects may be observed even for a single spin system after averaging over a time period much longer than all relevant time constants in the system. This ergodic nature of the individual spin dynamics has allowed the detection of single spins by means of fluorescence experiments.[4]

Specifically our interest towards the detection of spin noise arises from the possibility of studying individual molecules of molecular nanomagnets. Molecular nanomagnetism is a rapidly developing research field[5,6] whose aim, among others, is the design of molecules with desired magnetic properties. In this framework clusters possessing giant spins as high as $S= 51/2$ have been reported.[7] The detection of spin noise fluctuation is a promising technique to unveil the magnetic dynamics of these systems at the single molecule level. On the other hand, molecular nanomagnets might be particularly suited as benchmark tool for the development of new techniques to probe single spin dynamics.

The possibility of the detection of a single spin magnetic moment from its statistical polarization by means of Scanning Tunnelling Microscopy (STM) has been successfully proven more than a decade ago by one of the authors.[8] In the reported experiment the authors located an STM tip over a defect in oxidized silicon surface with a static magnetic field applied perpendicular to the surface plane.

Recent results[1,9] have proven that Electron Spin Noise induced coherent oscillations detected by STM (ESN-STM) can be observed not only in dangling bonds at silicon surfaces but also looking at organic radicals deposited on highly oriented pyrolytic graphite (HOPG) surfaces. By locating the STM tip over an assembly of organic radicals the authors were able to detect a peak for a molecular system with electron spin S=1/2. The same authors have demonstrated that the technique is able to detect side peaks located at the position of the organic radical hyperfine levels.[9]

In this paper we report on the verification of recent results on the detection of spin noise from paramagnetic molecules of BDPA (α,γ-Bisdiphenylene-β-phenylallyl) by Durkan and coworkers[1] and we focus on new results on a second paramagnetic specie, DPPH (1,1-Diphenyl-2-picrylhydrazyl), deposited on Au(111) surfaces. These two organic radicals molecules DPPH and BDPA are well known Electron Spin Resonance (ESR) standards and their properties in solutions have been studied through ESR and Electron Nuclear Double Resonance (ENDOR). It is relevant to this study that for both species the electron spin is substantially delocalised on the entire molecule.[10,11] This has implications on the type of electron spin dephasing process as explained in the discussion.

The difficulty of the experiment requires a thorough report on the technique and experimental apparatus we have built which will be given in section II. Experimental details will be given in section III while the obtained results will be discussed in sections IV. Further technical details on the apparatus are reported in the appendixes.

## II. IMPLEMENTATION OF ESN-STM EXPERIMENT

A major experimental difficulty encountered in the realization of the ESN-STM

experiments[1,12] lies in the fact that the signal level is quite low (it was estimated to be -120 dBm on a 50 Ω impedance line[1]) and inherently the signal to noise (S/N) ratio[1,9] amounts only to 4÷5. It is not clear whether this low intensity is entirely due to the intrinsic physics of the observed phenomenon or whether it might be improved by a superior RF recovery circuitry.

The dependence of the mean square uncertainty in the measurement of the spectral density *S*, under certain approximations,[13] is inversely dependent on the observation time *t*:

$$\frac{\langle \Delta S \rangle}{S^2} = \frac{1}{(RBW)t} \tag{1}$$

where *RBW* is the resolution bandwidth of the spectrum analyser. The use of the STM to probe a spin center limits the observation time to a few seconds due to the thermal drift of the STM tip. On the other hand the field heterogeneity in the x, y, z direction demands a frequency span of the measurements (typically from 10 to 20 MHz in our setup). These two experimental constraints are clearly in conflict with each other and create a major source of experimental difficulties in electron spin noise detected by STM. Furthermore unlikely other experiments on spin noise fluctuations in spin ensemble[3,4,14] the dependence of the spectral density on the spin dynamics in STM experiments is largely unknown. A number of theoretical models have been recently reported but they have not been experimentally verified.[12,15-25] However, these theories postulate an interaction between the electron spins of the paramagnetic site and the tunnelling electrons. Previous studies on spin noise,[26] have outlined that a near field $B_1$ generated by the ensemble of spins itself is responsible for the induction of a voltage into picking up coils. The importance of the near field in detecting single spins and small spin ensembles is also revealed by the recent work spin detection with other nanoprobes.[27-29] Herein we do not refer to a specific model and associated spectral density to analyze our data.

We have used a general protocol presented by one of the authors to detect the low level RF

signal from the STM tip.[13,30] A small oscillating magnetic field $\Delta B$ is added to the static magnetic field $B_0$. In this way the actual field at the sample is $B = B_0 + \Delta B \cos(\omega_m t + \phi)$ where $\omega_m = 2\pi\nu_m$ is the modulation frequency ($\nu_m$ in units of Hz), $\phi$ is the phase. The resulting signal will occur at a frequency modulated in time $\omega = \omega_0 + \Delta\omega \cos(\omega_m t + \phi)$ where $\omega_0$ is the unmodulated frequency and $\Delta\omega = 2\pi g \mu_B \Delta B / h$ is the frequency modulation intensity (here $\mu_B$ is the Bohr magneton, $g$ is the Landé factor of the paramagnet and $h$ is the Planck's constant). The Fourier transform of such a signal will result in a set of equally spaced sidebands with frequencies $\omega_0, \omega_0 \pm \omega_m, \ldots, \omega_0 \pm n\omega_m$.[31] The intensity of the $n$-th side band will be given by a $n$-th order Bessel function of the first kind, $J_n(m_\omega)$, where $m_\omega = \Delta\omega/\omega_m$ is the modulation index. The number of side bands is roughly given by $2m_\omega$ so that the total width of the spectrum is given by $2\Delta\omega$. If the modulation index is chosen as $m_\omega = \Delta\omega/\omega_m = 2$, the components of the Fourier spectrum $J_1$ will be maximised and inherently a Phase Sensitive Detection (PSD) will provide the highest sensitivity in the detection of the signal.[13]

Fig.1 shows the set up we have used to detect spin noise fluctuation. The video output of the spectrum analyser is fed into a lock in amplifier referenced at $\omega_m = 2\pi\nu_m$. The spectrum analyser is actually a superheterodyne detector with three mixers that convert the input signal frequency to a Intermediate Frequency (IF) and then filters the high frequency components through a band pass filter. The signal from the band pass filter is detected by an envelope detector. The envelope detector provides a low frequency signal proportional to the absolute value of the IF signal $V_{out} = \langle |V_{IF}| \rangle$. If the band pass filter has a resolution band width, $RBW$, of the same order of the modulation frequency ($RBW \approx \nu_m$) a PSD can be used. The lineshape of the lock-in amplifier output will be dependent on a number of parameters and in particular on the ratio between the sweep time ($SWT$) and the lock-in time constant $\tau_{PSD}$. When the latter is smaller than the 1% of $SWT$ and the intrinsic linewidth of the signal is larger than $RBW$ the output of the lock-in amplifier will provide a derivative line shape. In principle in this situation the linewidth of the signal measured by the lock-in amplifier correspond to the intrinsic one and is correlated with the spin longitudinal decay time $T_1$ and the phase memory time

$T_2$.[3]

However, alongside with previous literature on single spin detection,[4] our results (see infra) demonstrate that the assumption that $\omega_0$ is a constant over the time of measurement $T_m$ is incorrect if interaction of the electron spin with the surrounding is considered.

A more realistic picture is described by the following equation:

$$\omega = \omega_0(t,T_1) + \cos(\omega_m t + \phi(t,T_2)) \qquad (2)$$

where $\omega_0(t,T_1)$ and $\phi(t, T_2)$ are random functions of the time with characteristic correlation times that are not known. In particular, $\omega_0(t,T_1)$ is a function containing the dependence of the central frequency on the hyperfine interaction, and on the $g$- anisotropy; it depends on the longitudinal relaxation time $T_1$; $\phi(t, T_2)$ is the function containing the dependence of the phase on the transverse relaxation time, $T_2$.

While single spin detection by MRFM is sensitive to the spin flip in the z component of the spin vector,[28] our setup is sensitive to the x and y components of the spin vector precessing around the z axis. As a result $T_1$ will mostly contribute to determine the time scale of frequency jumps, whereas $T_2$ will determine the timescale of the shortest spin sensitive interaction between the paramagnetic molecule and the probe. For ESN-STM there is no simple way to correlate the measured parameters to the intrinsic spin dynamics time constants, $T_1$ and $T_2$.

Another important aspect of our detection apparatus is the amplification and the overall noise figure. We have designed and constructed an RF recovery circuitry that can detect a low level signal in a relatively wide band (50 ÷ 1500 MHz). There are two important duties that the recovery circuitry should accomplish: recover the maximum signal power available and amplifying the signal without a severe degradation of the S/N ratio.

In order to achieve the former, the best solution is to place the first amplification stage as close as possible to the signal source, the tunnelling junction. Indeed, part of the signal can be lost into connections with the ground provided by several parasitic capacitances and this loss should be minimized. Fig. 2 illustrates a schematic of the tunnelling junction with its parasitic capacitances where $C_T$ is the tunnel junction capacitance, $C_c$ the connection capacitance (due to the connection between the source and the first amplification stage) and $C_B$ is the STM tip-body capacitance. In this paper we consider the junction capacitance $C_T$ to be purely geometrical. The cable connecting the STM tip to the RF preamplifier has been chosen to minimise $C_C$. A perfect impedance matching between the cable and the STM tip termination over the entire frequency range of interest would not be possible. Since the RF preamplifier is located close to the STM tip, the amount of signal backscattered at the cable-tip junction will be minimised for signal frequencies corresponding to wavelength bigger than the STM tip length. This is indeed the case for ESN-STM.

The sensitivity, $S$, expressed in dBm, of a recovery circuit including a spectrum analyser (SA) (the detector) is given at room temperature by:[32]

$$S = -174 + 10\log(RBW) + NF_{System} \qquad (3)$$

where *RBW* is the spectrum analyser Resolution Band Width and $NF_{System}$ is the Noise Figure of the entire recovery circuit till the spectrum analyser. In the case that the sum of the pre-amplification total gain and its noise figure is larger than 39 dBm, the $NF_{System}$ value is expressed by the following formula:[32]

$$NF_{system} = G_{PA} - 2.5 dB \qquad (4)$$

where $G_{PA}$ is the gain in dB of the pre-amplification stage. The sensitivity, as already mentioned, is limited by the acquisition time. In non-digital filtering spectrum analyzer the acquisition time *SWT* critically depends on the *RBW* and the on the frequency range (*SPAN*) in which the spectrum analyser search for a signal:

$$SWT = \frac{k * SPAN}{RBW^2} \qquad (5)$$

where *k* is a constant.[32]

In cases in which the signal is very close to the noise level, video filtering is used to flatten the spectrum and make the signal more evident. In these cases the *SWT* depends also on the Video Band Width filter (*VBW*) as follows:

$$SWT = \frac{k * SPAN}{RBW * VBW} \qquad (6)$$

The minimum acceptable value for *SPAN* is given by the interval in which the signal frequency fluctuates for a given hyperfine level.[1,31] However the widest span in which the signal must be recovered is given by the magnetic field measurement precision. Our Spectrum Analyzer is able to perform a span of 20 MHz (corresponding for an electron spin to about 7 Gauss error bar in the measurement of the magnetic field) in 500 ms with *RBW*= 10 KHz. We have taken *RBW*=*VBW*=10 KHz as design parameter for the RF recovery circuit. Also we have constructed a home built RF preamplifier with a *NF*= 1dB in a frequency span of (50 ÷ 1500 MHz). The I/V converter is incorporated in the same case as the RF preamplifier.

The frequency response of the RF amplifier is flat within 1dB and the gain *G* is larger than 20dB in the frequency range of interest. Further details are reported in Table 1. In cascade to this amplifier we have connected a Minicircuits ZKL-1R5 amplifier capable of 40 dB gain over a band width of 1500 MHz. The total gain of the pre-amplification stage ranges from 55 to 60 dB. The noise figure $NF_{PA}$ is 4dB. The sum of these two values is larger than 39 dB therefore eq. 2 holds.[32] The $NF_{system}$ is then 1.5 dB. This means that the theoretical sensitivity of the recovery system for a *SPAN* value of 20 MHz, and a *RBW*=10 KHz varies between -127.5 and -132.5 dBm as computed from eq.1.

Filtering of the RF environmental noise has proven to be crucial to the success of the experiment reported below. For this reason all the cable connections to the vacuum chamber have been filtered. Piezo drivers, motors and Hall probe connections have been filtered at the feedthroughs of the vacuum chamber by means of home built passive RF filters. These filters feature 70 dB of attenuation above 10 MHz. RF spurious noise is filtered also at the tunnelling current and bias voltage connections. RF harmonics due to digital equipment have been eliminated.

The connection between the second RF amplifier and the spectrum analyser has been decoupled from ground with a home built SMA ground de-coupler. It is a passive element with an attenuation larger than 2 dB below 2 GHz. This was necessary to reduce some low frequency noise that entered the STM through the spectrum analyser.

We have tested the sensitivity of our experimental set up by simulating a real experiment: a RF frequency signal is input into the tunnel junction; the entire experimental apparatus is set as in a real experiment with all the electronics equipment on (including field modulation); both unmodulated and modulated signals were input to compare spectrum analyzer detection only and spectrum analyzer detection combined with PSD detection.[3]

Table 2 reports the value of the sensitivity measured in the two cases at different frequencies. The PSD detection is more sensitive than spectrum analyzer stand alone detection. Also the derivative line shape of a PSD recorded spectra is more evident to the experimentalist while looking for a low level signal than the sharp shape peak close to the noise level provided by spectrum analyzer detection.

Another aspect which was crucial for the success of our ESN-STM experiments was the implementation of a software that can handle in real time both STM operations and ESN-STM. A key point of this software is its ability to scan acquiring an STM image, stop scanning at the operator command, position the tip over a target, modify tunnelling current and bias voltage, trigger on the spectrum analyser, acquire an ESN-STM spectra from the lock-in and restart scanning, reiterating this cycle every time the operator decides. In this way the software allows for acquiring many spectra and the STM topography associated with it and makes possible statistical analysis of the results.

### III. EXPERIMENTAL DETAILS

Samples for ESN-STM were prepared by immersing a flame annealed Au(111) 150 nm thick film evaporated on mica into $CH_2Cl_2$ solutions of 1,1-Diphenyl-2-picrylhydrazyl (DPPH) and α,γ-Bisdiphenylene-β-phenylallyl (BDPA), respectively (purchased from Sigma Aldrich. Inc.). The solutions concentration has been varied from 0.1 to 0.01 mM and the exposure time from 1 minute to 15 minutes in order to evaluate different degree of coverage and to optimize the deposition quality. The sample was then rinsed 3,4 times for 10-30 s into pure $CH_2Cl_2$, and dried under nitrogen fluxing.

STM imaging and ESN-STM were performed under ambient conditions. The vacuum chamber is sealed and is used to insulate the experimental set up from environmental RF spurious noise. The ESN-STM measurements reported in the following have been achieved under the application of an AC and a DC magnetic field. The AC field magnitude was 10 mG corresponding to a modulation index $m_\omega=2$ for a modulation frequency $v_m=$ 15 KHz. The AC field was measured using the AC option in the Sypris gaussimeter applying field of 0.1-0.5 Gauss and extrapolating the value of the voltage amplitude to achieve a field of 10 mG. Once the sample was mounted on the sample holder all components (namely $x$, $y$ and $z$) of the magnetic field on the sample were measured carefully. The field was measured[33] with a precision of 2-4 G depending on the size of the sample mounted on the sample holder.

In order to distinguish between ESN-STM spectra and the environmental noise at the same frequency, over 600 spectra were taken with the tip slightly retracted and out of tunnelling. The entire experimental apparatus was set up as for measurements performed with the tip in tunnelling. (Peaks in the spectral density of the tunnelling current can arise from spontaneous voltage fluctuations at the RF amplifier input as well as at the modulating coils, this circumstance needs to be ruled out.[26] After acquiring the spectrum from the lock-in amplifier in these conditions, the three largest peaks per spectrum were analyzed. The peak to peak amplitude $P_{kk}$, divided by the standard deviation $\sigma$, is the parameter that was used for the comparison.

CW-ESR spectra were acquired using a Bruker Elexsys E500 spectrometer working at X-Band (ν~ 9.4 GHz) equipped with a SHQ cavity. Ultrathin film samples for investigation with this technique were obtained by incubating Au(111) flame annealed slides for 4 hours in $CH_2Cl_2$ 0.1 mM solutions of the radicals.

## IV. RESULTS AND DISCUSSION

### 1. STM

Fig. 3a and 3b show typical images observed for DPPH and BDPA molecules deposited on Au(111) surfaces. As the geometrical size of the smallest white spots corresponds to DPPH and BDPA molecular size within experimental error, we attribute them to DPPH and BDPA single molecules. Larger white spots indicates the formation of agglomerates of two or few molecules. ESN-STM signal reported herein were always obtained from single molecules.

Molecules deposited by spontaneous adsorption from diluted solutions did not form agglomerates with a large vertical size as found in the drop casting method. STM imaging was achieved only for tunnel currents below 30 pA. The obtained STM images of the molecules are elongated along the scanning direction. We attribute this effect to an internal reorientation of the molecule caused by the tunnelling current. Similar effects were reported recently.[34]

A relevant issue is whether the difficulty of STM imaging is related to molecular diffusion processes on Au(111). To the best of our knowledge the diffusion properties of DPPH and BDPA on Au(111) are not known. However several studies have been conducted by Ultra High Vacuum Variable Temperature STM to determine migration energy and diffusion constants of large adsorbates on metal surfaces.[35] The hopping rate of the molecule on the surface depends upon the molecular surface interaction, the surface geometry and also the surface coverage with molecular adsorbates. Both DPPH and BDPA are molecules containing several aromatic rings that can interact with noble metals via π interaction. In the case of molecules like $C_{60}$ the migration energy on Au(111) is 1500 meV[35] and as a result diffusion is basically precluded at room temperature. Other molecules like PVBA (4-*trans*-2-(pyrid-4-yl-vinyl) benzoic acid, a molecule with two phenyl rings) deposited on Pd(110), have an hopping rate of about 0.01Hz at room temperature.[35] On the basis of data collected on other molecules we can then conclude that on average the molecule under the STM tip will not diffuse away during spectroscopy at the coverage we used. STM images acquired after a spectra was taken on a molecule demonstrate that this did not change its position.

Fig 4c and 4d shows ordered agglomerates of BDPA molecules. These are formed after 24-36 hours from deposition. Once the molecules are laterally confined on the surface the STM image quality improves as long as the tunnel current is confined below 50 pA. These data are similar to what observed before from one of the authors for molecular complexes,[36] in which long range diffusion was ruled out.

## 2. Continuous wave ESR

Continuous wave ESR (CW-ESR) measurements on BDPA and DPPH radicals proved that these molecules retain their paramagnetic character on the Au(111) surface (Fig. 4). An ESR signal is observed at g=2.005(5) for BDPA with $\Delta H_{pp}$=2.5 G. It is interesting to compare the spectral appearance of this ultrathin film with those obtained respectively for BDPA solution samples (dilute paramagnetic centres) and drop cast BDPA films on Au(111) (paramagnetic centres densely packed). Spectra from solution appeared to be severely broadened, with $\Delta H_{pp}$ = 7.0 G and with gaussian lineshape, independent of the solution concentration (in a range between 1 μM and 100 μM). On the other hand drop cast thick films showed a signal with an approximately lorentzian lineshape and $\Delta H_{pp}$ = 2.5 Gauss. This different behaviour is evidenced in the inset of the left part of Fig. 4 where the absorption curves of drop cast samples and ultrathin films are reported, showing that the intensity of the latter is partially observed on the tail of the resonance line, indicating that exchange narrowing processes are less effective than in the case of drop cast samples.[37-39] This suggests that in the latter samples molecules are close enough to interact between themselves whereas a somehow better isolation of each molecule is obtained in ultrathin films. This is likely to happen also for sub-monolayer coverage films as the ones used to achieve ESN-STM spectra.

The ESR spectrum of DPPH containing sample shows a single line with $\Delta H_{pp}$ = 2.8 G (right part of Fig. 4). No hyperfine structure is observed, in contrast with solution spectrum, showing the expected five lines pattern due to the hyperfine coupling with two almost equivalent $^{14}$N I=1 nuclei (coupling with $^{1}$H being unresolved). This is again attributed to exchange narrowing processes. The comparison of thin film samples with drop cast ones does not evidence, in this case, major differences in line shape and/or linewidth, suggesting an easier formation of aggregates for DPPH than for BDPA.

In conclusion, CW-ESR results prove unambiguously that the paramagnetic character of the molecules is retained after deposition on gold. Further they suggest that, at least for BDPA, the interactions between paramagnetic molecules of drop cast films are strongly reduced in ultra thin films that were used for ESN-STM measurements. Even if this can not be considered as a conclusive evidence of the presence of isolated molecules it clearly demonstrates that the aggregates which may be present are composed of a very small amount of molecules. This is in agreement with what is inferred from STM images.

**3. ESN-STM**.

The statistical distribution of over 1800 noise fluctuations extracted from spectra obtained with the tip slightly retracted and out of tunnelling is shown in Fig. 5. The analysis of the peak evidence that the value $P_{kk}/\sigma$ for these spurious signals due to environmental noise was never larger than 6.7.

In the following we report only ESN-STM peaks with a value $P_{kk}/\sigma \geq 7$. The statistical distribution reported in the inset of Fig. 5 shows the distribution of the $P_{kk}/\sigma$ for successful ESN-STM measurements.

Over 3300 ESN-STM spectra were measured on samples as described above. Only 0.5% of them resulted in ESN-STM spectra. Fig.6a shows the ESN-STM peak at 650.5 MHz (232.3 G) measured on the DPPH molecule evidenced by a circle in Fig.6b. The peak has a derivative shape as reported in previous work.[13] In this specific case the tunnel current during spectroscopy was raised to 0.6 nA while scanning was achieved at 30 pA. When a peak appeared in the spectrum the position of the tip was always localized on a single molecule. As only the 0.5% of the spectra taken resulted in ESN-STM spectra, we cannot rule out the possibility that also the observed ESN-STM signal from a single molecule might be somehow affected by the nearby presence of an agglomerate of 2-3 molecules.

Fig. 7a reports the dependence of the peak position in the ESN-STM spectrum at different value of the applied DC magnetic field. The theoretical linear dependence of the Larmor frequency with varying the magnetic field, which is the most important indicator for spin detection,[28] is verified.

Fig. 7 b, c, d illustrate the shape and the magnitude of the peaks at 651.7, 539.1 and 427.3 MHz, corresponding to field of 232.75, 192.5 and 152.6 G, respectively. At each field the experiments were done in a range of frequencies of about 20 MHz (the vertical bars in Fig. 6a and 7d, one single scan was never larger than 10 MHz). Nevertheless the signals were observed at the right frequency with a precision of 3 MHz. This shows that signals are not observed at frequencies different from the Larmor frequency.

Finally we have reproduced Durkan results on BDPA. The tunnelling current during ESN-STM spectroscopy is always confined in the range 0.3 to 0.6 nA, less than half the value reported in Durkan's experiments.[1,9]

Fig. 8b reports the ESN-STM spectrum detected on the BDPA molecule highlighted in Fig. 8a. This spectrum was measured with a lock-in amplifier sensitivity of 200μV and has to be compared with the spectrum in Fig. 9c where the same lock-in sensitivity for measurements on DPPH has been used. The frequency of the peaks measured is consistent with the measured magnetic field as illustrated in Fig. 8d. The uncertainty in this case is obviously higher than the graph reported in Fig.7a as only the sample position on the plane is varied, while the magnet $z$ position is always constant(effect of B inhomogeneity in the x and y direction) .

An analysis of the bandwidth of the signal observed for DPPH reveals a non monotonic dependence of the bandwidth on the frequency. Peaks found at the lowest frequency (150 G region) have a width bigger than 400 KHz whereas peaks at 540 and 650 MHz (190 and 234 G) regions have band widths that spans from 100 to 300 KHz. Comparisons between DPPH and BDPA peak widths for fields in the 234 G region shows values that are comprised in the range 100 to 300 KHz.

## 4. DISCUSSION

The intensity of the ESN-STM peaks measured for both DPPH and BDPA is lower than that reported by Durkan[1] for BDPA even if it has to be noted that for those experiments no statistical data on success rate and amplitude distribution were reported. The magnitude of the PSD peak detected by one of the authors with the same technique was also higher.[13,30] We believe that this is the result of a complex relationship between the spin dynamics of individual paramagnetic centres, their electronic structure and our detection apparatus. We will analyze this problem in detail in the following.

The ruling concept herein is that the spectrum analyzer relies on a bandwidth filter that is swept across a certain span of frequencies. A signal can be reliably detected only if it is present at a given frequency for the entire duration of the sweep. If the signal has a transient nature with an intrinsic instability in frequency, the spectrum analyzer might fail to detect it. In the following we discuss a number of causes that might make the signal transient and therefore decrease the probability that it can be captured at the right time. This is particularly true if the sweep rate is slowed down and the time scale of the transient signal duration is small compared to this rate.[40]

In our experimental set up the sweep speed has been decreased to enable the use of a lock-in amplifier and our ESN - STM spectra are acquired for time of up to 6 s. This is a time scale in which the molecule can undergo intramolecular transition several times.[34] As a result the molecule might be in a state that does not produce a spin signal at the moment in which the bandwidth filter is tuned at the Larmor Frequency.

Moreover the hyperfine interactions in BDPA and DPPH reduce the amplitude of the detectable signal by spreading the actual band in which the signal can be found over a range of about 50 MHz.[10,11] When the STM tip is brought over a molecule to detect the signal the receiver is tuned to the Larmor frequency of the central $m_I$ level within a band of 10 MHz. There is however a possibility that during the measurement the level populated is a different $m_I$ one. If the time of observation is long enough and the bandwidth of the receiver covers the entire hyperfine frequency range $\Delta\omega_I$, then the ESR spectrum of a single spin can be extracted.[4] In other words, hyperfine interactions can lead to an ESN-STM transient signal depending on the nuclear spin flip rate that is strongly enhanced in the presence of a electron spin[41,42] and depends on the electron spin longitudinal decay time $T_1$. Because the unpaired electron in both DPPH and BDPA are delocalised over the entire molecule[10,11] several nuclear spins (the most relevant ones are the $^1$H atoms with I=1/2 and, only for DPPH, the $^{14}$N atoms with I=1) might flip during the measurement leading to a possible change in the energy of the electron spin state occupied. This situation does not apply to the Pb centers that were previously investigated by one of the authors,[31] in fact the natural abundance of $^{29}$Si (I=1/2) is low (4%) and most of the Pb centres on $SiO_2$ surface do not present hyperfine coupling.[43,44,45] It has to be noted here that, according to the results of CW-ESR, on a macroscopic scale the hyperfine interaction is largely averaged out due to exchange narrowing effects. If we attribute the observed reduction in ESN-STM signal intensity to hyperfine coupling we may then speculate that while CW-ESR spectra are dominated by signal of small molecular aggregates, the molecules we have selected for investigation through ESN-STM are much more isolated than their average on a macroscopic scale and are not affected by exchange coupling.

The frequency at which the signal can be detected might also be affected by local electric field oscillations. However this is true for Pb centres on SiO$_2$ rather than in organic radicals as reported in literature.[46] For organic radicals the effect of *g*- anisotropy will produce a fluctuation of the spin noise frequency within a band of a 1-2 MHz. This can be calculated by taking into account some early publications on solid DPPH.[47,48] While these fluctuations fall in the bandwidth of our receiver they are larger than the signal width and are driven by intra-molecular motion which might be triggered by the tunnelling current.[34] The *g*- anisotropy is also present on Pb centres studied before but the alignment of the spin centre with the direction of the magnetic field in a rigid lattice can not change.

As a further point it has to be noted that $T_1$ is longer for semiconductors than for paramagnetic molecules like BDPA and DPPH, for which it also depends on the molecular environment.[41,42,49] In general crystalline organic radicals or concentrate solutions show $T_1=T_2$ due to exchange narrowing, however in a more dilute environment $T_1$ becomes bigger than $T_2$ and dependent on the temperature.[50] Early studies[51-53] of BDPA in solution show a $T_1$ of the order of 1μs while other studies[11] report 0.5 μs. The value of $T_1$ for DPPH appears to be slightly slower but on the same order of magnitude.[54,55] This seems to indicate that $T_1$ might be the parameter that ultimately determine the rate at which the ESN signal frequency jumps to a different value.

Other factors that might account for the discrepancy in the signal intensity with respect Durkan[1] experiments might be the following:

(i) The value of the tunnelling current during our ESN-STM experiment is always in the range 0.3 – 0.6 nA as opposed to the 1.4 nA used in the Durkan's one. The lower value of the tunnelling current was necessary to avoid damaging the tip and the surface at the location of the tip. In this way we could take several ESN-STM spectra for every scanned image.

(ii) The geometrical capacitance at the tip-sample junction can affect the S/N ratio. Our tips were not chemically etched and therefore may have larger geometrical capacitance.

(iii) The molecules studied here were deposited on Au(111) and not HOPG surface. Even if it is still not clear how the nature of the surface may affect the signal, internal molecular motion are much probably modified by a different electronic and vibronic environment[57,58].

**VII CONCLUSIONS**

We have detected spin noise oscillations from two distinct paramagnetic species deposited at the Au(111) surface. These results prove and extend previous experiments by Durkan[1] but also point out the difficulty of this experiment. We have employed a method of detection designed by Manassen[31] and improved it by the construction of a new equipment. We give a detailed explanation of our experimental setup to allow interested scientists to have a good starting point to develop their own instrumentation.

The detection of magnetic resonances through noise has been so far confined to a fundamental interest playground. However, the demand for characterisation tools at the single molecular level and protocols for spin read out in Quantum Computing give impulse to further develop the method for possible future applications. Single spin detection at room temperature however, poses strong experimental challenges. Our results suggest that future work will have to focus on a RF detection apparatus capable to measure fluctuations on the time scale in which they are generated in a single molecule (detection of transient signals). In other words it will be required to detect frequency fluctuations on a large bandwidth and to extract signal from the noise in a large detection bandwidth. This is a truly outstanding experimental challenge. Results might also be improved by an appropriate choice of molecular systems. e.g. paramagnetic species with a long $T_1$ and reduced hyperfine interactions. Also a thorough Ultra High Vacuum STM study of the candidate molecular specie will be necessary to help selecting molecules that have low diffusion rates on the surface and undergo as small as possible intramolecular rearrangement during the spectroscopy time.

A further issue is the understanding of the coupling mechanism between the transversal components of the spin vector and the detection apparatus. So far theories have focused on describing how tunnelling electrons might couple to the noise while the possibility of near field effects[2,13,26-29] has been ruled out. However, if the coupling between the spin noise detection system and the STM tip were due to a near field effect or a type of electrodynamics pick up[56] it would be possible to design experiments where the tunnelling current does not perturb the molecular adsorbates inherently improving the probability to detect the signal.

**ACKNOWLEDGMENTS**

(*Further acknowledgements will appear in the accepted version*) L. Lenci and C. Ascoli from the IPCF, Pisa, Italy, are acknowledged for providing the magnet and for fruitful discussions. F.Fradin , from ANL, US, is acknowledged for a revision of the manuscript. R. Sergo and A. Grunden, from Elettra, Trieste, Italy , are acknowledged for their technical support.

Financial support from the Italian MIUR, FIRB and FISR projects, EC HPRI-CT-2000-40022 SENTINEL and MRTN CT-2003 504880 "QuEMolNa" and NoE "Magmanet" are also acknowledged . Partial support from the US-Israel binational foundation is also acknowledged.

* To whom correspondence should be addressed. Currently at Material Science Division, Argonne National Laboratory, 9700 S. Cass Avenue Argonne, IL 60439 E-mail: pmessina@anl.gov

**APPENDIX A**

The STM head is placed inside a vacuum chamber. The chamber provides an easy load for samples with adsorbed molecules on their surface. The pressure reaches $10^{-5}$ Torr within one hour. The STM head is suspended on stainless steel springs (vibration damping).

The microscope body is divided into two blocks. The bottom part allocates an X-Y coarse translation stage and the piezo-tube. The X-Y coarse positioning stage was added to the system to allow the operator to change the position of the tip over a large area. It is particularly useful to relocate the tip after performing ESN-STM spectroscopy that may locally damage the molecular films (tunneling currents are typically from $I_t$=0.3 to 1.5 nA ). The sample is mounted on the bottom of the upper part of the main body . The coarse approach motion is done by an electrical motor coupled with the feedback controlled Z movement of the piezo-tube.

The STM head top body has a hole allowing the permanent magnet to slide forward and backwards with respect to the sample. The measurement of the magnetic field is accomplished through the Hall probe, Sypris 5080, mounted on top of the sample holder . The probe resolution is 0.1 Gauss (~0.28 MHz for a paramagnet with an electronic Landé factor g=2.00), the accuracy is 1% of the reading in the range of 0.1÷30 000 Gauss. The magnet (NdFeB) is driven forward and backwards by a software controlled electrical motor. A precise home built mechanical positioning device drives the Hall probe over the sample.

The damping system was proven to work effectively. The RF cabling connection does not significantly alter the transfer function of the damping stage. The RF low noise, low level broad band preamplifier is placed close to the STM tip. The decoupling circuit and the DC amplifier are both located inside the case with the RF amplifier. The tunnelling current is coupled to the DC amplifier through a set of integrated inductances. These components have a twofold action: on one hand they prevent RF signal to be lost in the DC part of the circuit; on the other hand they reduce the parasitic capacitance towards the ground seen by the amplifier. The former is required as the maximum amount of RF signal must be transferred to the RF amplifier. The latter is due to the fact that parasitic capacitance seen by the DC amplifier increases the noise at the input. As the DC amplifier is set to detect very low currents, the capacitance must be reduced as much as possible.

The I/V converter was designed to satisfy two requirements: i) imaging molecules at low tunnelling current (typically 1 to 10 pA) ii) performing ESN-STM spectroscopy on molecules anchored on the surface ($I_t= 0.3 \div 1.5$ nA).

The two major difficulties encountered in the design are the limiting extra Johnson noise coming from the RF amplifier and the noise current at the non-inverting terminal due to the stray capacitance.[59,60] The measured noise level for this home built part is 0.6 pA RMS. The feedback loop controller can stabilise a current as low as 1 pA.

Because of the AC field applied in time modulated ESN-STM experiments, a four poles home built filter (-3dB point at 10 KHz) was added to the I/V converter output to avoid STM feedback oscillations.

To verify the ability of the I/V converter to work with the operating RF recovery system self-assembled monolayers of hexa-decanethiol on Au(111) surfaces were prepared by dipping Au (evaporated on muscovite mica) slides into 1mM solution of thiol in ethanol, as reported in literature.[61,62] Good STM images were achieved in a pressure of $10^{-5}$ torr at 2pA with the RF amplifier switched off and at 20 pA with the RF amplifier switched on.

**APPENDIX B**

The perfect matching of the junction impedance and the RF input of the RF amplifier at any frequency was not possible. To simulate different STM tip lengths, we built several assemblies consisting of a plate and a coax cable with external shield piled out at different lengths. We also embedded these assemblies into Teflon pieces to simulate different RF paths for grounding the shield of the coax cable. The inner conductor coaxial cable was brought to different distances from the base plate. This prototype assembly is not measured in tunnelling conditions, however this is not relevant here as the overall impedance of the sample STM tip depends upon its geometrical characteristics.

The impedance measured in different configurations was always much higher than the amplifier input impedance of 50 Ω. Fig. 9a shows an example impedance measurement of one of these assemblies. As in our STM design the coaxial cable between the STM tip and the RF amplifier is about 10 cm long the signal arriving to the amplifier is already adapted to a 50 Ω line. For this reason the input circuit to the amplifier was designed at 50 Ω. The amplifier circuit layout is outlined in Fig. 9d. The core component is the MGA-62563. The 6.8 nH inductor in the input connector of the MGA-62563 provides the impedance matching between the DC-AC splitter and the input of the MGA-62563. The MGA-62563 is a low noise RF amplifier in E-pHEMET GaAs technology. Linearity is excellent. The external resistance of 1.5 KΩ blocks the amplifier polarization current to 40 mA. The characteristics of the amplifier inserted into the circuits allows a noise figure lower than 1dB.

We also report in Fig. 9b, the frequency response of the ground decouple inserted between the RF amplifier and the spectrum analyzer.

Figure Captions:

**Fig. 1.** Detection scheme for ESR STM. The signal is recovered from the STM tunnelling current and is split into the DC and AC part. The AC part is then amplified and detected by a spectrum analyzer (SA). The video out put of the SA is connected to the input of a lock-in amplifier which detects the component of the signal in phase with the AC magnetic field generated by the magnet's coil. The part within the dashed line illustrates the function of the SA: the signal is sampled and mixed with a reference signal produced within the analyzer. The mixed signal is filtered and amplified. The width of the filter (RBW) sets the frequency resolution with which a spectral feature can be resolved. As explained in the text the RBW and the lock in integration time determine the time of a single span. The picture shows also the extensive filtering to spurious noise that has been implemented on this set up.

**Fig. 2** The tunnel junction and the cable connections to the RF preamplifier are schematically shown as a current generator with a parallel impedance. The impedance has a resistive part and several capacitances. The origin of the several capacitances is illustrated in the drawing on the bottom right. The junction capacitance $C_j$ accounts for both the geometrical capacitance and the quantum capacitance (see text and references). In order to transfer the maximum signal to the RF amplifier ($Z_{LOAD}$ in the drawing on the bottom left) the sum of the capacitances must be minimized. This is achieved by shortening the distance between the STM tip and the RF preamplifier.

**Fig. 3 a)** STM image (10×10 nm$^2$) of DPPH molecules deposited on Au(111). Tunnelling current I$_t$= 10 pA, Bias Voltage BV= 0.1 V. **b)** STM image (15×15 nm$^2$) of BDPA molecules deposited on Au(111). I$_t$= 50 pA, BV= 0.1 V. **c)** STM image (50×50 nm$^2$) of BDPA conglomerates on Au(111) achieved 36 hours from deposition. I$_t$= 50 pA, BV= 0.1 V **d)** STM image (20×20 nm$^2$) of BDPA conglomerates on Au(111) achieved 36 hours from deposition.. I$_t$= 50 pA, BV= 0.1 V.

**Fig. 4** Left: Room temperature CW-ESR spectra of BDPA (molecule sketched in the upper right inset) as ultrathin film sample (a), drop casted sample (b), and dichloromethane 0.1 mM solution (c). In the upper left inset the comparison between the integrated spectra of samples (a) and (b) is reported.

**Fig. 4** right: Room temperature CW-ESR spectra of DPPH (molecule sketched in the upper right inset) as ultrathin film sample (a) and dichloromethane 0.1 mM solution (b)

**Fig. 5** Distribution of the amplitudes of the three largest noise fluctuations in over 630 spectra taken with the tip slightly retracted and out of tunneling. The value of the peak is normalised with respect the standard deviation of each spectra acquired. **Inset**: distribution of ESN-STM peaks showing a value Pkk/σ ≥7.

**Fig. 6 a)** ESN-STM spectrum of DPPH deposited on Au(111) showing a peak at 651.5 MHz (232.7 G). SPAN = 649-652 MHz, BW = VBW = 30 KHz, SWT = 6 sec. The parameters during ESN-STM measurement were: tunneling current: 0.6 nA, bias voltage 0.3 V. AC field modulation frequency and intensity: 15 KHz, 10 mG. Lock in sensitivity 1 mV, time constant 10 ms. **b)** STM image (10×10 nm$^2$) of the molecule in a)..Experimental conditions V during STM imaging: I$_t$= 30 pA, BV= 0.3.

**Fig. 7 a)** Position of ESN-STM peak measured at different values of the DC magnetic field applied. The horizontal error bars represent the error in the magnetic field measurement. The vertical bar indicates the frequency range in which the ESN-STM signal was searched. The line is calculated from the formula $\nu = \frac{1}{2}g\mu_B B$, in units of MHz and Gauss: $\nu(MHz) = 2.8 \frac{MHz}{Gauss} \times B(Gauss)$.

**b)** ESN-STM spectrum showing a peak at 427.13 MHz (152.54 G). SPAN= 425-432 MHz, BW=VBW=30 KHz. SWT=6 sec. Tunneling current during spectroscopy 0.3 nA Bias voltage during spectroscopy 0.3 V. AC field modulation 15 KHz, 10 mG. Lock in sensitivity 1 mV, time constant 10 ms.

**c)** ESN-STM spectrum showing a peak at 539.1 MHz (192.5 G). SPAN= 536-540 MHz, BW=VBW=30 KHz. SWT=6 sec. Tunneling current during spectroscopy 0.3 nA. Bias voltage during spectroscopy 0.3 V. AC field modulation 15 KHz, 10 mG. Lock in sensitivity 1 mV, time constant 10 ms.

**d)** ESN-STM spectrum showing a peak at 651.68 MHz (232.7 G). SPAN= 648-658 MHz, BW=VBW=30 KHz. SWT=6 sec. Tunneling current during spectroscopy 0.6 nA. Bias voltage during spectroscopy 0.3 V. AC field modulation 15 KHz, 10 mG. Lock in sensitivity 1 mV, time constant 10 ms.

**Fig. 8 a)** STM image (7×7 nm$^2$) of a BDPA molecule on which the ESN-STM spectrum reported in b) is measured...$I_t$= 30 pA, BV= 0.3 V during STM imaging.

**b)** ESN-STM spectrum showing a peak at 664.8 MHz (237.4 G).. SPAN= 660-670 MHz, BW=VBW=30 KHz. SWT=6 sec. Tunneling current during spectroscopy 0.3 nA. Bias voltage during spectroscopy 0.3 V. AC field modulation 15 KHz, 10 mG. Lock in sensitivity 200 μV, time constant 10 ms.

**c)** ESN-STM spectrum taken on a different BDPA molecule showing a peak at 658.3 MHz. SPAN= 660-670 MHz, BW=VBW=30 KHz. SWT=6 sec. Tunneling current during spectroscopy 0.3 nA. Bias voltage during spectroscopy 0.3 V. AC field modulation 15 KHz, 10 mG. Lock in sensitivity 1 mV, time constant 10 ms.

**d)** ESN-STM peaks were detected at the expected frequency. The horizontal error bar represents the magnetic field range measured over the sample surface. The vertical error bar indicates the frequency range in which the ESN-STM signal was searched. The frequency range was divided in two o three sub-range that were investigated separately.

**Fig.9 a)** Impedance measurement of a cable assembly a base plate that simulates the Tip-sample geometrical assembly.

**b)** Frequency response of the Ground decoupler. Insertion loss is less than 1 dB. **c)** Frequency response of the DC port of the AC-DC splitter**. d)**

Tables

| Property | Value |
|---|---|
| Maximum Gain | 22.6 dB @ 200 MHz |
| Band width with Gain ≥ 10 dB<br>Band width with Gain ≥ 15 dB<br>Band width with Gain ≥ 20 dB | 32.6 to 4100 MHz<br>45.5 to 2700 MHz<br>73.8 to 1200 MHz |
| Minimum Signal detected | -140 dBm @ 500 MHz |
| Noise level reported to the input<br>Input impedance | -145 dBm @ 500 MHz<br>50 Ω |
| Output impedance<br>Power supply | 50 Ω<br>20 to 30 V |
| Current *demand* | < 30 mA |

**Tab. 1** Major features of the home built low level low noise RF amplifier. This amplifier is the first stage of the recovery circuitry.

| Frequency (MHZ) | SA Sensitivity (dBm) | SA sensitivity (dBm) | Lock-in sensitivity (dBm) | Lock-in sensitivity (dBm) |
|---|---|---|---|---|
| | SWT= 4 s SPAN= 6 MHz | SWT= 4 s SPAN= 3 MHz | SWT= 4 s SPAN= 6 MHz | SWT= 4 s SPAN= 3 MHz |
| 100 | -95 | -93 | -99 | -98 |
| 200 | -109 | -109 | -116 | -117 |
| 300 | -131 | -130 | -136 | -139 |
| 400 | -119 | -119 | -122 | -122 |
| 500 | -130 | -130 | -136 | -136 |
| 600 | -137 | -137 | -145 | -147 |
| 700 | -128 | -125 | -138 | -135 |
| 800 | -114 | -116 | -135 | -137 |
| 900 | -134 | -133 | -140 | -137 |
| 1000 | -138 | -139 | -144 | -148 |

**Tab. 2** Comparisons between the sensitivity of the recovery circuit when spectrum analyzer is used as a final detector and when a Lock-in is used instead

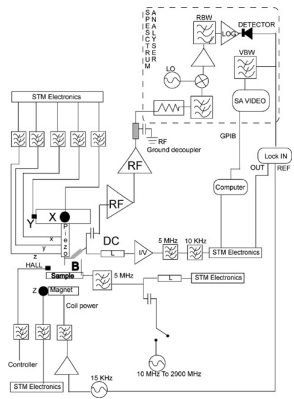

Fig.1

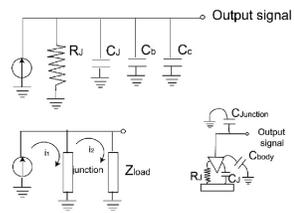

Fig.2

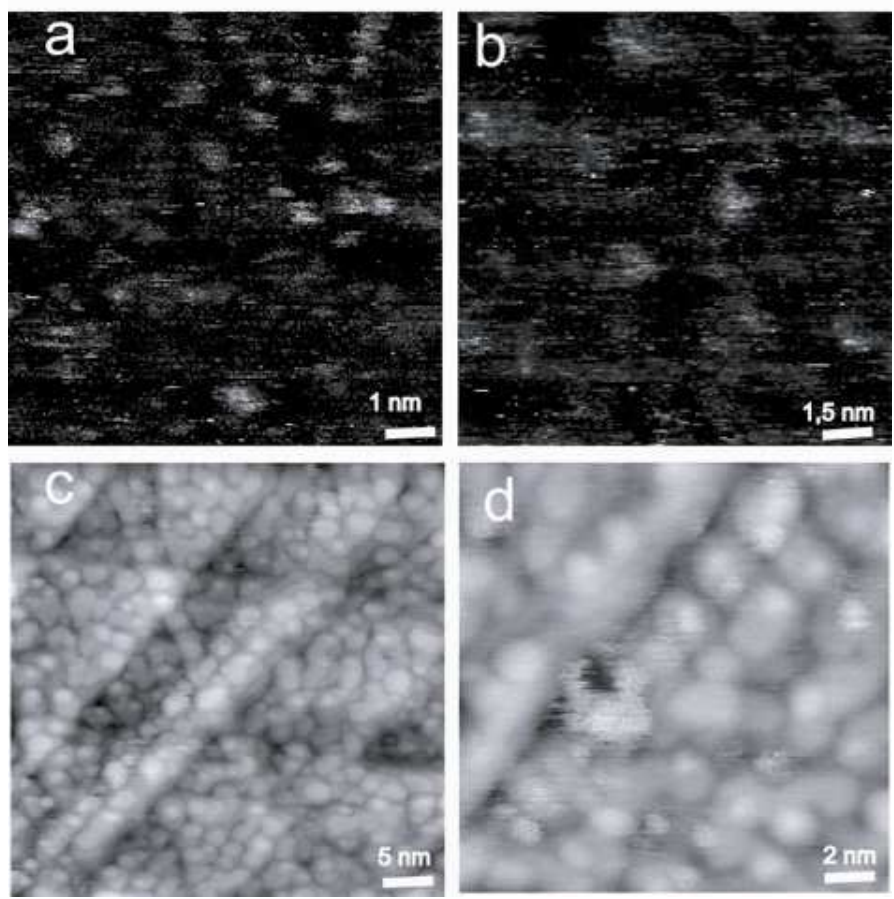

Fig.3

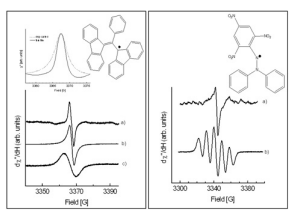

Fig.4

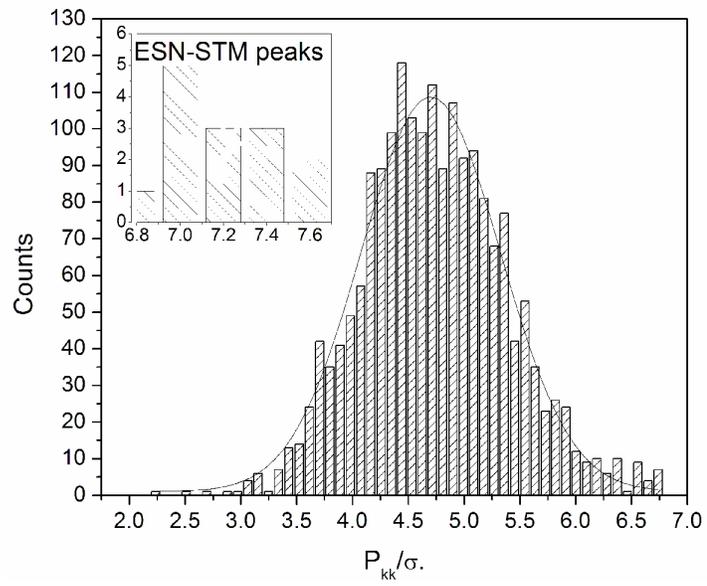

Fig.5

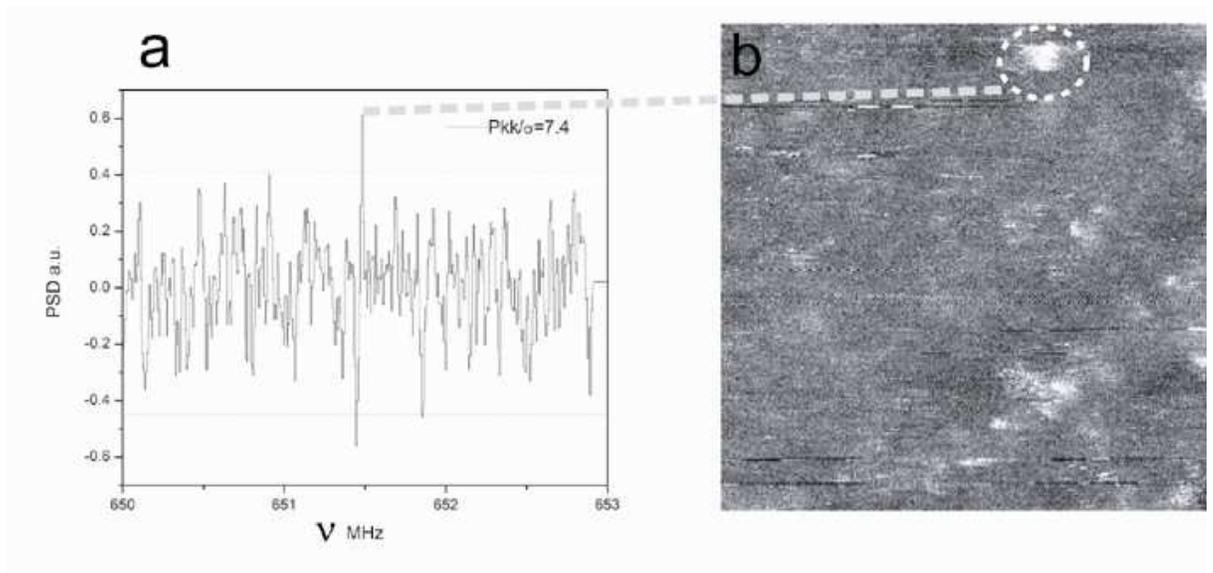

Fig.6

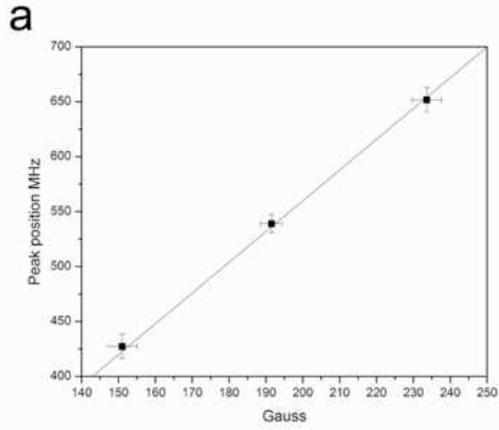
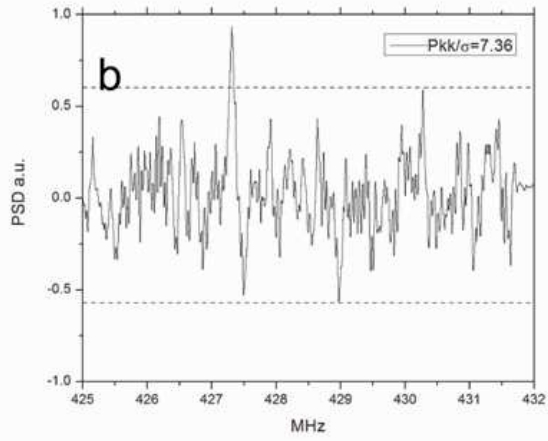
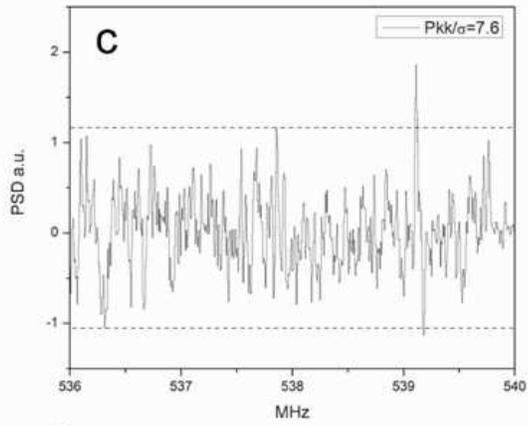
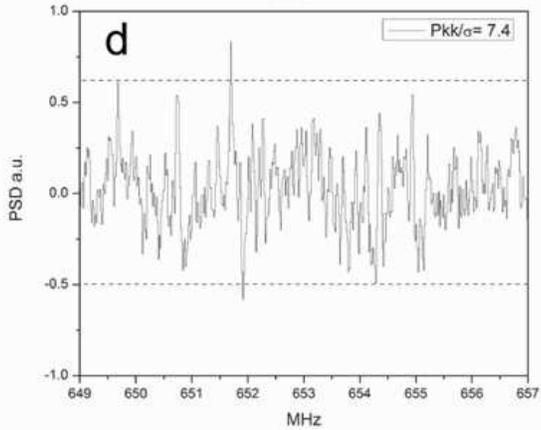

Fig.7

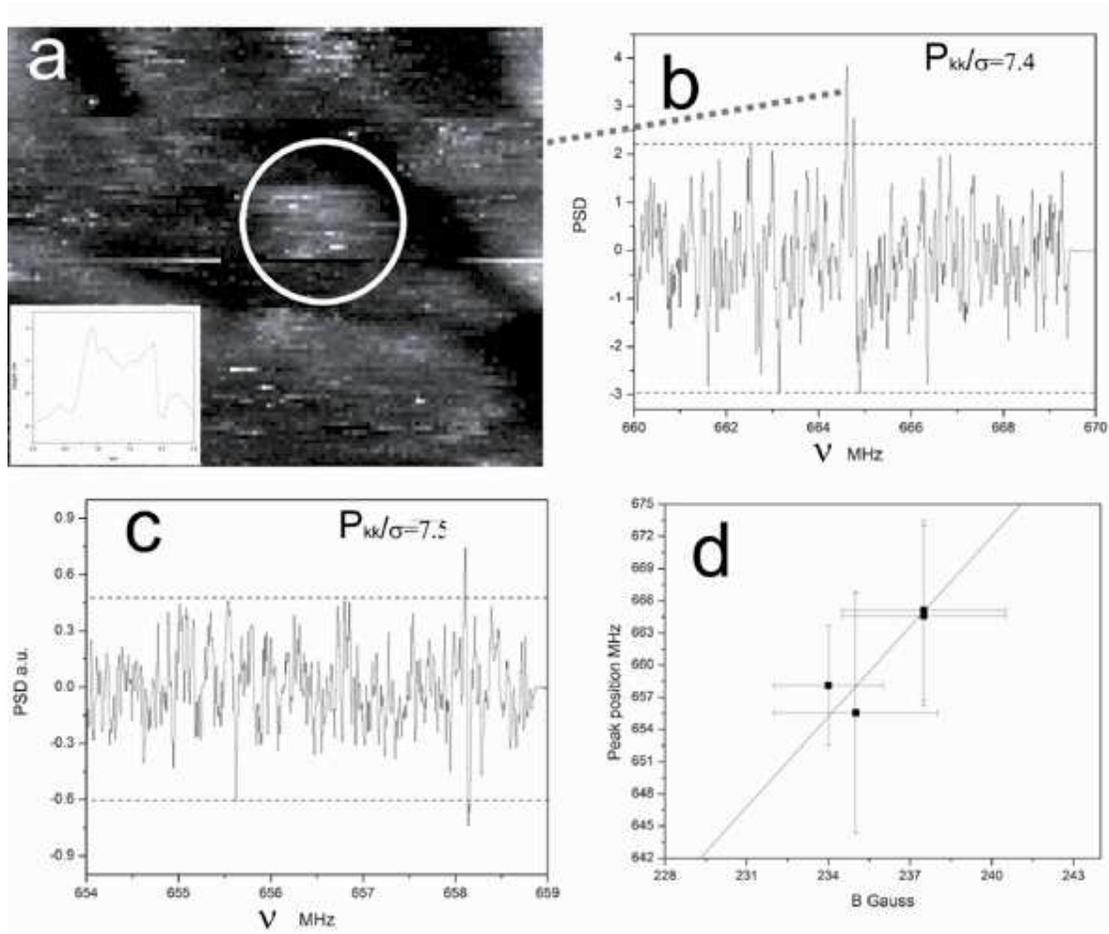

Fig.8

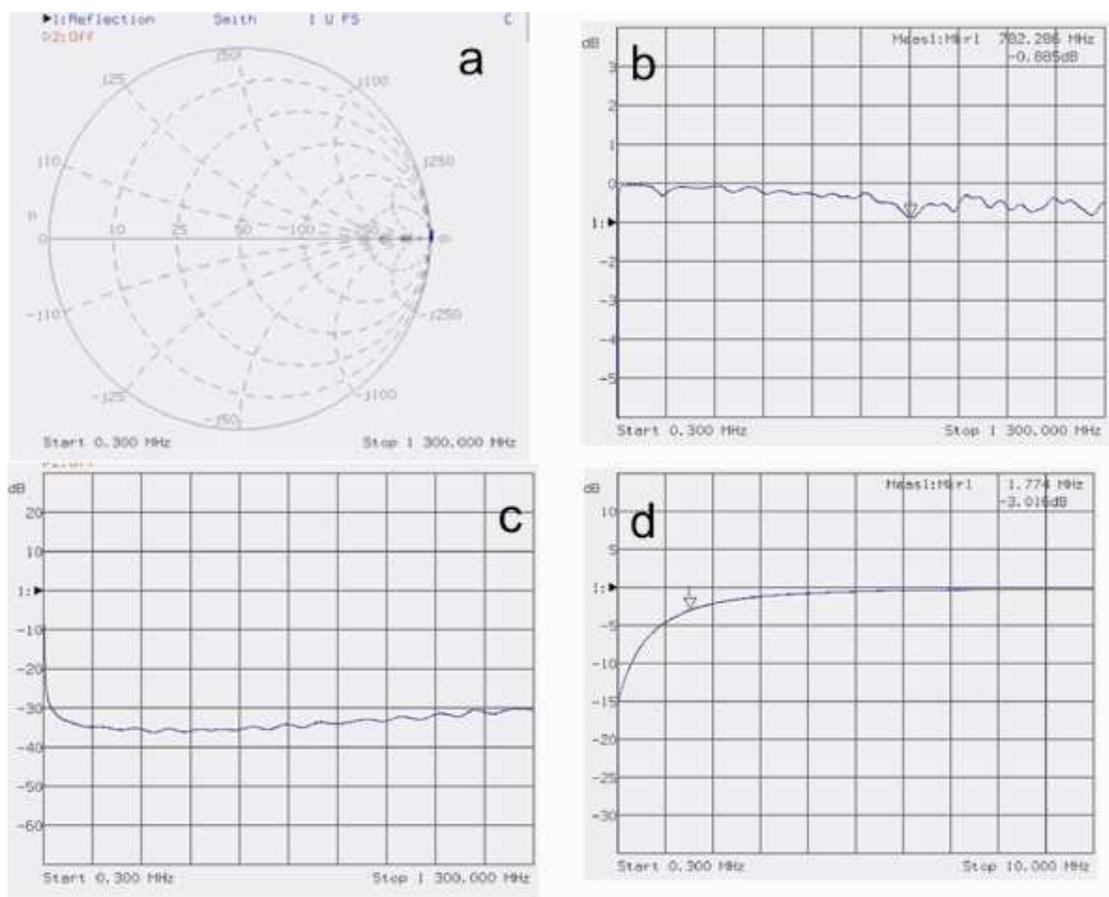
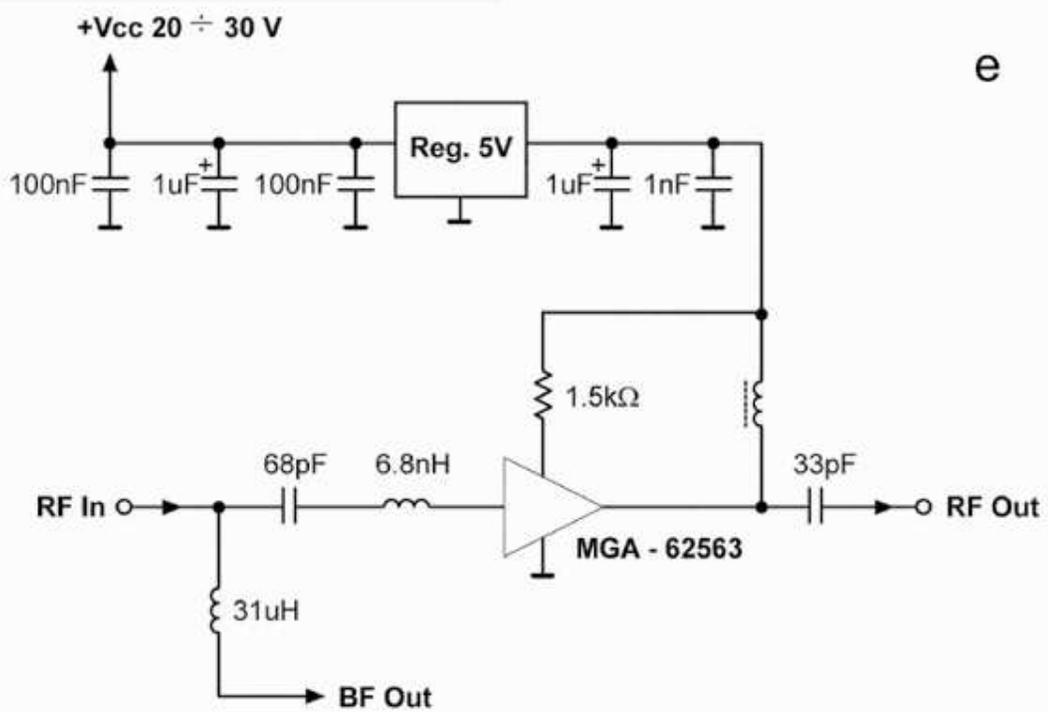

Fig.9